\documentclass[prl,twocolumn]{revtex4}
\usepackage{graphicx}

\begin{document}

\title{Pressure-driven flow of solid helium}

\author{James Day}
\author{John Beamish}

\affiliation{Department of Physics, University of Alberta,
Edmonton, Alberta, Canada, T6G 2J1}

\date{\today}

\begin{abstract}
The recent torsional oscillator results of Kim and Chan suggest a
supersolid phase transition in solid $^4$He.  We have used a
piezoelectrically driven diaphragm to study the flow of solid
helium through an array of capillaries.  Our measurements showed
no indication of low temperature flow, placing stringent
restrictions on supersolid flow in response to a pressure
difference.  The average flow speed at low temperatures was less
than 1.2x10$^{-14}$ m/s, corresponding to a supersolid velocity at
least 7 orders of magnitude smaller than the critical velocities
inferred from the torsional oscillator measurements.
\end{abstract}

\maketitle

Recent experiments by Kim and Chan\cite{Kim04-225,Kim04-1941}
showed that solid helium decouples from a torsional oscillator at
temperatures below about 0.2 K. In liquid $^4$He, such decoupling
reflects the non-classical rotational inertia (NCRI) associated
with superfluidity and these experiments suggest that $^4$He also
exhibits \textquotedblleft supersolidity". The possibility of
supersolidity in helium has been discussed for many
years\cite{Andreev69-1107,Leggett70-1543,Meisel92-121} but
previous experimental
searches\cite{Greywall77-1291,Bishop81-2844,Bonfait89-1997} were
unsuccessful. Following Kim and Chan's experiments, a number of
papers have discussed the possible microscopic origins of
supersolidity\cite{Prokofev05-155302,Burovski05-165301,Galli05-140506,Saslow05-092502,Anderson05-1164,Ceperley05-155303,Dai05-132504,Bernu05-1462}
and the properties that such a state might
exhibit\cite{Saslow05-092502,Dorsey05a}. However, there is not yet
a consensus on whether supersolidity can occur in a defect-free
crystal and further experiments are needed to establish whether
solid helium displays any of the other unusual properties
associated with superfluidity. We recently\cite{Day05-035301} used
a capacitive method to look for pressure-driven flow of solid
helium confined in the pores of vycor glass, but saw no evidence
of superflow at temperatures down to 30 mK, nor has supersolidity
been seen in recent ultrasonic experiments in
vycor\cite{Kobayashi05a}. In this Letter, we report measurements
of DC and low frequency AC flow of solid $^4$He through an array
of glass capillaries. Near the melting temperature, applying a
pressure difference caused solid helium to flow through the
capillaries, but the rate decreased with temperature; below about
1 K no flow was detected. Our experiments extended to 35 mK, well
into the temperature range where Kim and Chan observed NCRI, and
used isotopically pure $^4$He. Our results place stringent limits
on possible pressure-induced supersolid flow.

The essential results of the torsional oscillator measurements
were similar for $^4$He confined in the nanometer pores of vycor
glass\cite{Kim04-225} and for bulk $^4$He\cite{Kim04-1941}. Each
showed a gradual transition at T$_c$ $\approx$ 0.2 K with about
1$\%$ of the helium (the \textquotedblleft supersolid fraction"
$\rho$$_s$/$\rho$) decoupling at the lowest temperatures and
amplitudes. The decoupling was smaller at large oscillation
amplitudes, suggesting a supersolid critical velocity v$_c$ $\sim$
10 $\mu$m/s in both systems. The similarities support the
interpretation that NCRI is an intrinsic property of solid helium
rather than, for example, occurring in liquid layer at pore
surfaces. The measurements in vycor revealed a remarkable
sensitivity to $^3$He impurities; concentrations as low as 10 ppm
significantly reduced the NCRI. In bulk $^4$He, $\rho$$_s$ was
shown to vary with pressure, going through a maximum around 55
bar. However, T$_c$ was nearly pressure independent.

Early suggestions\cite{Andreev69-1107} that $^4$He could exhibit
supersolidity were based on the idea that quantum solids might
contain \textquotedblleft zero point vacancies" (ZPV) which would
bose condense and produce supersolidity. However, both
measurements\cite{Meisel92-121} and
calculations\cite{Pederiva97-5909} indicate that vacancies in
helium have an activation energy of at least 15 K, with no
evidence of vacancies at zero temperature. Nonetheless, direct
comparisons of density to lattice constants from x-ray
measurements can only rule out ZPV at the 0.1$\%$ level and
Anderson recently suggested\cite{Anderson05-1164} that solid
helium may be incommensurate, with vacancies in a highly
correlated ground state. Even in the absence of ZPV, Leggett
showed\cite{Leggett70-1543} that atomic exchange could lead to
supersolidity.  He estimated $\rho$$_s$/$\rho$ $\lesssim$
10$^{-4}$ for $^4$He but subsequent
calculations\cite{Galli05-140506,Saslow05-092502} have predicted
supersolid fractions ranging from 10$^{-5}$ to 1$\%$. However,
recent path integral monte carlo
calculations\cite{Ceperley05-155303} found that exchange
frequencies decrease exponentially with ring length and thus that
supersolidity is not expected in a perfect, commensurate $^4$He
crystal. The conclusion that supersolidity in $^4$He must involve
defects is supported by very general path integral
arguments\cite{Prokofev05-155302,Burovski05-165301}.
Vacancy-interstitial pairs (VIP) may be needed for
supersolidity\cite{Prokofev05-155302,Burovski05-165301,Galli05-140506,
Dai05-132504} but interstitials in $^4$He have large activation
energies (around 48 K\cite{Ceperley05-155303}) and VIP appear to
be strongly bound\cite{Bernu05-1462} and so cannot transport mass
or produce supersolidity. Extended defects such as dislocations,
stacking faults or grain boundaries may be essential.

While there is not yet a consensus on the microscopic origin of
supersolidity, the similarity between the NCRI seen for $^4$He in
the pores of vycor and for bulk $^4$He constrains models. For
example, it is difficult to imagine mechanisms involving grain
boundaries that would not be affected by confinement in nm pores.
Recent calculations\cite{Khairallah05-185301} for $^4$He in
vycor-like pores provide evidence of a mobile liquid-like layer
near the pore surface where the superfluid response might
originate. This could be related to the NCRI's sensitivity to
$^3$He, since impurities would preferentially go to the
delocalized layer and disrupt superfluidity, but it would not
explain the bulk helium results.

The long-standing interest in quantum crystals inspired a number
of earlier searches for supersolidity in $^4$He, although many of
them did not reach the temperature range where Kim and Chan
observed NCRI. One that did extend to 25 mK saw no decoupling from
a torsional oscillator\cite{Bishop81-2844}, leading the authors to
conclude that either the transition temperature was below 25 mK or
else the supersolid density or the critical velocity was very
small ($\rho$$_s$/$\rho$ $<$ 5x10$^{-6}$ or v$_c$ $<$ 5 $\mu$m/s).
There have also been attempts to look for flow of solid helium in
capillaries but pressure differences of order 1 bar did not
produce measurable flow\cite{Greywall77-1291} down to 30 mK, nor
was flow seen in a subsequent U-tube
experiment\cite{Bonfait89-1997} which extended to 4 mK. These
measurements put similar limits on possible superflow in bulk
helium ($\frac{\rho_s}{\rho}$v$_c$ $\lesssim$ 2x10$^{-11}$ m/s)
and our recent experiments\cite{Day05-035301} put a comparable
limit on pressure-induced flow solid of $^4$He in the pores of
vycor ($\frac{\rho_s}{\rho}$v$_c$ $\lesssim$ 1.5x10$^{-11}$ m/s).
One group of
experiments\cite{Lengua90-251,Ho97-409,Goodkind02-095301} that did
show unusual behavior involved ultrasound and heat pulses. The
interpretation was complicated but, like Kim and Chan's
observation of NCRI, the results were sensitive to $^3$He
impurities at the ppm level. These were the only experiments to
date which used isotopically pure $^4$He.

Other than the torsional oscillator experiments, there have not
yet been direct observations of supersolid behavior either in bulk
or in small pores. However, the small critical velocities and the
sensitivity to $^3$He impurities may affect DC flow or other
properties even more strongly than the torsional oscillator
measurements. Also, solids have properties not shared by liquids
(e.g. a lattice with shear rigidity) and a supersolid may not
exhibit all of the effects we associate with superfluidity (e.g.
superleaks, persistent currents, thermomechanical effects,
quantized vortices, second sound, etc.). Below we describe a set
of experiments to look for one such property of solid $^4$He:
superflow in response to pressure. We applied small pressure
differences (3 to 100 mbar) at low temperatures (down to 35 mK)
and used both isotopically pure $^4$He ($^3$He concentration $<$
0.002 ppm\cite{bureauofmines}) and $^4$He with the natural
isotopic composition (typically 0.3 ppm $^3$He). We made both DC
and low frequency AC (below 1 Hz) measurements, but did not see
any evidence of flow below about 1 K.

Our beryllium copper cell consisted of two cylindrical chambers
connected by a \textquotedblleft superleak" of about 36,000
parallel glass capillaries (25 microns in diameter) which were
fused into a 3 mm thick  \textquotedblleft glass capillary array"
(GCA\cite{collimatedholes}) with an open cross-sectional area A =
0.18 cm$^2$. The outer wall of the larger chamber (diameter 25 mm,
height $\approx$ 1 mm, volume V$_1$ = 0.49 cm$^3$) included a
flexible diaphragm which could be moved with an external PZT
piezoelectric actuator\cite{pzt} to compress the helium. The
smaller chamber (diameter 7 mm, height 0.3 mm, V$_2$ = 0.01
cm$^3$) included a capacitive pressure gauge which, when used with
a 1 kHz automatic bridge (Andeen-Hagerling 2550 A) had a
resolution and stability better than 0.2 mbar. If helium moves a
distance dx through the capillaries, the resulting pressure change
is dP = $\frac{A}{\kappa V_2}$dx , where $\kappa$ is the helium's
compressibility, so we typically could detect a 0.3 nm
displacement of solid $^4$He through the GCA. The cell, which had
a total volume (including the GCA channels and fill line)
V$_{total}$ = 0.79 cm$^3$, was mounted on the mixing chamber of a
dilution refrigerator. Temperatures were measured with a germanium
thermometer, with a $^{60}$Co nuclear orientation thermometer for
calibration below 50 mK.

We started by filling and pressurizing the cell at 4.2 K, using a
room temperature gauge to calibrate our capacitive pressure gauge.
We calibrated our PZT actuator and diaphragm in the liquid phase
at 1.95 K and 36.4 bar, just below the melting curve. The bottom
set of data in Fig. 1 shows the pressure response (right axis)
when the full voltage (150 VDC) was applied to the actuator. As
expected, the pressure increased immediately (within the few
seconds the capacitance bridge took to respond) and returned to
its original value when the diaphragm was released after about
half an hour. The pressure change due to the compression was
$\Delta$ P$_{liquid}$ $\approx$ 84 mbar. Using the liquid's
compressibility ($\kappa_{liquid}$ = 3.6x10$^{-3}$ bar$^{-1}$),
gives a volume change $\Delta$V/V$_{total}$ $\approx$ 0.03$\%$,
corresponding to a diaphragm deflection of about 1 $\mu$m.

Crystals were grown using the blocked capillary, constant volume
technique. We started with liquid at high pressure and monitored
the cell pressure as it was cooled. At a pressure of 61.7 bar,
freezing began at 2.60 K and was complete at a final pressure of
37.1 bar. Annealing the solid near its melting temperature
eliminated the pressure gradients created during freezing and
produced a sharp melting onset (at T$_m$ = 1.96 K) characteristic
of a uniform density crystal. Our initial experiments used $^4$He
with the natural isotopic composition and were consistent with the
measurements shown in this paper which were made using
isotopically pure $^4$He.

Our basic flow measurement was made at temperatures below T$_m$ by
quickly (over about 5 seconds) applying a DC voltage to the PZT
actuator to squeeze the solid $^4$He, thus increasing the pressure
in the large chamber. In contrast to the case where the cell
contained liquid, the solid helium may flow through the GCA
channels slowly, or not at all, so the pressures in the two
chambers may not equilibrate. However, even without flow, some
pressure is transmitted to the second chamber, since a pressure
difference will cause the GCA plate separating the chambers to
flex elastically. This small deflection appears as an immediate
pressure step in the other chamber. Any subsequent flow through
the channels will further increase the pressure, but more slowly.

\begin{figure}
\includegraphics[width=\linewidth]{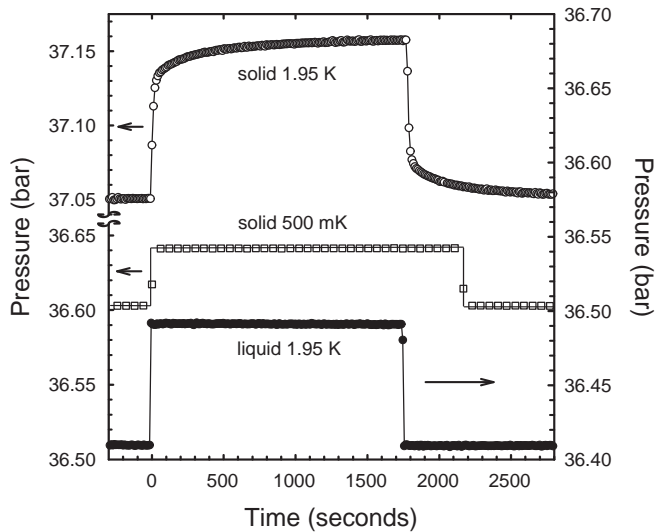}
\caption{Pressure response to \textquotedblleft squeezes". Lower
curve (solid circles): liquid $^4$He at 1.95 K, 36.4 bar. Middle
curve (open squares): solid $^4$He at 500 mK, 36.6 bar. Upper
curve (open circles): solid $^4$He near melting at 1.95 K, 37.1
bar.  Lines are guides to the eye. Note the different pressure
axes.} \label{fig:squeezes}
\end{figure}

The upper two sets of data in Fig.~\ref{fig:squeezes} show the
response to a pressure step when the cell contains solid helium.
At 0.5 K (middle curve) the pressure in the second chamber
immediately changed by about 38 mbar, corresponding to the GCA
flexing by about 30 nm.  Above about half the melting temperature,
this initial jump was followed by a slower, temperature-dependent
change due to flow. The top curve in Fig.~\ref{fig:squeezes} shows
the response at 1.95 K, very close to melting. After the initial
jump, the pressure continued to increase due to flow of solid
through the channels and relaxation of the GCA, but stabilized
within about half an hour. The total increase of 105 mbar is
slightly larger than the corresponding change with liquid helium,
as expected given the solid's smaller compressibility
($\kappa_{solid}$ $\approx$ 3.1x10$^{-3}$ bar$^{-1}$), and
indicates that, near melting, flow through the channels can
maintain pressure equilibrium between the two chambers. For all
three sets of data, we confirmed the linearity of the response,
i.e. the pressure changes were proportional to the voltage applied
to the diaphragm actuator.

\begin{figure}
\includegraphics[width=\linewidth]{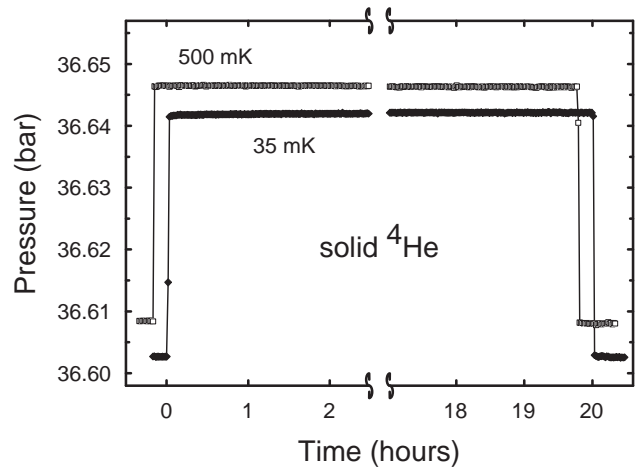}
\caption{Solid $^4$He response at 500 mK (upper curve, open
symbols) and 35 mK (lower curve, solid symbols). Lines are guides
to the eye and the curves are offset for clarity. Note the time
scale, which is much longer than in Fig.~\ref{fig:squeezes}.}
\label{fig:lowT}
\end{figure}

The most interesting question is whether solid helium will flow
through the capillaries in the temperature range where Kim and
Chan saw decoupling. Fig.~\ref{fig:lowT} compares the pressure
response at 35 mK to that at 500 mK. They are essentially
identical, with no indication of flow over a period of about 20
hours. The rate of pressure change is

\begin{equation}
\frac{dP}{dt} = \frac{A \bar{\text v}}{\kappa_{solid} V_2} <
\frac{0.5 mbar}{20 hours}\label{pressurechange}
\end{equation}

\noindent giving a limit on the average flow velocity

\begin{equation}
\bar{\text v} = \frac{\rho_s}{\rho}\text v_c < 1.2x10^{-14} m/s.
\label{averagevelocity}
\end{equation}

\begin{figure}
\includegraphics[width=\linewidth]{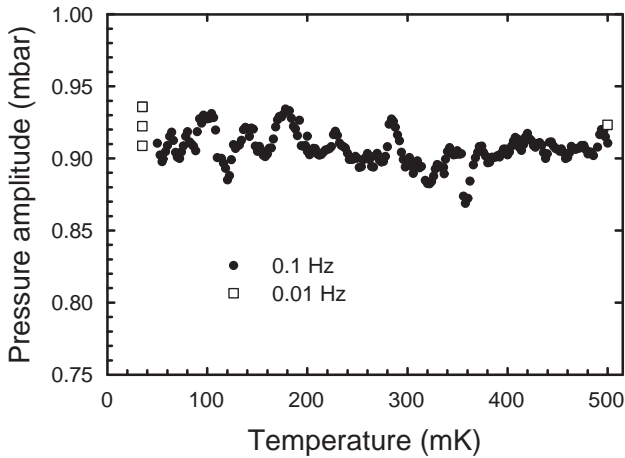}
\caption{AC pressure response in solid $^4$He at low temperatures.
Solid symbols are taken at 0.1 Hz during cooling. Open squares at
35 and 500 mK were taken at 0.01 Hz.} \label{fig:ac}
\end{figure}

We also made low frequency AC measurements using the piezoelectric
actuator to produce smaller pressure oscillations ($\pm$ 4 V,
corresponding to $\pm$ 3 mbar). The pressure was measured using a
manual capacitance bridge (General Radio 1615-A operating at 10
kHz) with an analog lock-in amplifier, and the AC response was
monitored with a digital lock-in (Stanford Research SR830 DSP). At
0.5 K the amplitude of the pressure oscillations was independent
of frequency up to about 1 Hz, as expected since the GCA can flex
very rapidly. Close to melting, the frequency dependence was more
complicated since, as Fig. 1 shows, solid can flow through the
capillaries even on a time scale of a few seconds. We looked for
AC flow at low temperatures by cooling the cell below 0.5 K.
Fig.~\ref{fig:ac} shows the amplitude of the pressure oscillations
at a frequency of 0.1 Hz. It also shows 0.01 Hz data at 35 mK and
at 0.5 K, illustrating the frequency independence over this
temperature range. The resolution is better than for DC flow and
the pressure amplitude is constant within $\pm$ 0.02 mbar, with no
evidence of temperature dependence that could be attributed to the
onset of flow through the capillaries. Sample heating limited
these measurements to frequencies below 1 Hz so we were not able
to make direct comparison to Kim and Chan's torsional oscillator
measurements at 1 kHz.

For a supersolid fraction $\frac{\rho_s}{\rho}$ = 1$\%$, our DC
flow limit~(\ref{averagevelocity}) implies v$_c$ $<$
1.2x10$^{-12}$m/s, seven orders of magnitude smaller than the
critical velocity inferred from Kim and Chan's torsional
oscillator measurements and more than a thousand times smaller
than the limits set by previous flow
experiments\cite{Greywall77-1291,Bonfait89-1997}. Flow in solids
often involves dislocations or grain boundaries, which can be
immobilized by small concentrations of impurities. Our
measurements using isotopically pure $^4$He were essentially
identical to our initial results with natural $^4$He so the
absence of pressure-induced superflow is not due to impurity
pinning of such defects. There has also been a
suggestion\cite{Dash05-235301} that a surface melted layer could
allow solid helium in a torsional oscillator to slip, providing an
alternative, non-supersolid explanation of the bulk $^4$He
decoupling. Our measurements appear to rule out such behavior at
low temperatures, although it may occur near melting.

The torsional oscillator results were also consistent with the
displacement, rather than the velocity, being limited to a
critical value. We can put limits on possible displacements of the
solid helium at low temperatures from the data in
Figs.~\ref{fig:lowT} and ~\ref{fig:ac}. Since the pressure jumps
at 35 and 500 mK agree within 1 mbar, the corresponding
displacements cannot differ by more than 2 nm. Our AC measurements
are less sensitive to flow, but more sensitive to displacements,
and rule out movements of solid helium through the capillaries
larger than 0.03 nm. If we again assume that only a 1$\%$
supersolid fraction moves, this would imply supersolid
displacements less than 3 nm, comparable to the amplitude of Kim
and Chan's torsional oscillator at their critical velocity (for
their 1 kHz oscillator, v$_c$ $\sim$ 10 $\mu$m/s corresponds to an
amplitude $\frac{v_c}{\omega}$ $\sim$ 2 nm).

These experiments show that static or low frequency pressure
differences do not produce either superflow or unusual
displacements at low temperatures in solid $^4$He. If the helium
forms a supersolid, then its flow properties must be quite
different from those of a superfluid, in which the chemical
potential difference created by a pressure change would cause
superflow.

This work was supported by the Natural Sciences and Engineering
Research Council of Canada.

\bibliography{bulk}

\end{document}